\documentclass{emulateapj}
\usepackage{natbib}

\shorttitle{X-ray Absorption Toward Abell 2029 due to a Foreground Spiral}
\shortauthors{Clarke et al.}

\begin{document}

\def\HI{\mbox{H\,{\sc i}}}
\def\kms{km~s$^{-1}$}
\def\hal{H$\alpha$}
\def\lsun{\mbox{$L_\odot$}}
\def\msun{\mbox{$M_\odot$}}

\title{Soft X-ray Absorption due to a Foreground Edge-On Spiral Galaxy Toward the Core of Abell 2029}

\author{T. E. Clarke\altaffilmark{1},
Juan M. Uson\altaffilmark{2},
C. L. Sarazin\altaffilmark{1},
and E. L. Blanton\altaffilmark{1,3}}

\altaffiltext{1}{Department of Astronomy, University of Virginia,
P. O. Box 3818, Charlottesville, VA 22903-0818, USA;
tclarke@virginia.edu, sarazin@virginia.edu, eblanton@virginia.edu}

\altaffiltext{2}{National Radio Astronomy Observatory, 520 Edgemont Road,
Charlottesville, VA 22903, USA;
juson@nrao.edu}

\altaffiltext{3}{Chandra Fellow}

\begin{abstract}
We have detected an X-ray absorption feature against the core of the
galaxy cluster Abell 2029 ($z=0.0767$) which we identify with the
foreground galaxy UZC J151054.6+054313 ($z=0.0221$).  Optical
observations ($B$, $V$, $R$, and $I$) indicate that it is an Scd
galaxy seen nearly edge-on at an inclination of $87^\circ \pm
3^\circ$.  \HI\ observations give a rotation velocity of 108 \kms\ and
an atomic hydrogen mass of $M_{HI} = 3.1 \times 10^9 d_{90}^2$~\msun,
where $d_{90}$ is the distance to the galaxy in units of 90 Mpc.
X-ray spectral fits to the $Chandra$ absorption feature yield a
hydrogen column density of $( 2.0 \pm 0.4) \times 10^{21}$ cm$^{-2}$
assuming solar abundances.  If the absorber is uniformly distributed
over the disk of the galaxy, the implied hydrogen mass is $M_H = (6.2
\pm 1.2) \times 10^8 d_{90}^2 \, $ \msun.  Since the absorbing gas in
the galaxy is probably concentrated to the center of the galaxy and
the middle of the disk, this is a lower limit to the total hydrogen
mass.  On the other hand, the absorption measurements imply that the
dark matter in UZC J151054.6+054313 is not distributed in a relatively
uniform diffuse gas.
\end{abstract}


\section{Introduction}

In this paper we report on the serendipitous detection of a foreground
edge-on galaxy seen in absorption against the X-ray core of a cooling
flow galaxy cluster. The cluster, Abell 2029, is a nearby ($z=0.0767$)
cluster which contains a central cD galaxy whose diffuse light extends
up to 850 kpc \citep*{ubk91}. It has been (re)classified by
\citet{d78} as richness class 4.4. X-ray studies of the cluster show
that it is a very relaxed system, and is one of the most luminous
galaxy clusters [$L_x (2 - 10\,{\rm keV}) = 1.1 \times 10^{45}$\ ergs
s$^{-1}$]. We have examined the archival $Chandra$ image of the
cluster to study the structure of the thermal intracluster medium
(ICM) in the core of the cluster and the interaction of the central
radio source with the ICM (T.\ E.\ Clarke et al., in preparation,
hereafter Paper II). While analyzing the $Chandra$ image, we
discovered a linear absorption feature roughly 1\farcm5 south of the
cluster core.

This feature is due to the galaxy UZC J151054.6+054313 (hereafter UZC
J151054), which is a beautiful example of a late-type spiral galaxy
seen (nearly) edge-on in the foreground of the dense cluster
Abell~2029.  It has a distinctive bulge component and a rather thin
disk with a disk-to-bulge ratio of $\sim 6$, consistent with the
parameters of an Sb/Sc galaxy (Figure~\ref{fig:optic_xray}a).
\citet{d81} measured redshifts of 31 galaxies toward Abell~2029 and
found UZC J151054 to be in the foreground of the cluster, at a
redshift of $z=0.0221$, while the cluster is at a redshift of
$z=0.0767$. The linear feature is due to photoelectric absorption of
soft X-rays in the disk of the spiral galaxy which is seen against the
diffuse thermal intracluster medium. In \S~\ref{obs}, we discuss the
X-ray, optical, and radio observations and reductions for this
galaxy. The optical and \HI\ details of the system are given in
\S~\ref{opt_prop} and \S~\ref{HI_prop},
respectively. \S~\ref{X-ray_prop} presents the X-ray absorption due to
the foreground spiral, and in \S~\ref{dis} we discuss the results of
these observations.

Throughout this paper, we adopt a luminosity distance of 90 Mpc for
UZC~J151054. We express the physical properties in terms of this
distance, and scale quantities with $d_{90}$. This distance is
consistent with a pure Hubble Flow and WMAP cosmology \citep{wmap},
which gives $D_L$=93.2 Mpc.  This is also in agreement with the
Tully-Fisher distance of $D_L\sim$ 83 Mpc (see \S~\ref{HI_prop}
below). At a distance of 90 Mpc, the scale is 0.44 kpc/arcsec.

\section{Observations and Data Reduction}\label{obs}

\subsection{X-Ray}

The central region of Abell 2029 was observed with the $Chandra$ X-ray
Observatory on 2000 April 12 for a total time of 19.8 ksec (OBSID
891). The observations were centered on the back-illuminated ACIS-S3
chip. We present a detailed description of the data reductions in
Paper II and only provide a brief summary here. The data were analyzed
using CIAO version 2.3 and CALDB 2.18. The events were filtered for
energy (0.3--10 keV) and grade (ASCA 0,2,3,4, and 6). The background
light-curve was determined from the S1 chip and showed no strong
flares, and we removed only 128 seconds of data due to high
background.

The background for the observation was determined from the blank sky
background files collected by M. Markevitch\footnote{See
http://cxc.harvard.edu/contrib/maxim/acisbg/COOKBOOK}.  For the
purposes of studying the absorption due to the foreground galaxy, it
is the soft X-ray background in the 0.3--1 keV band which is most
important.  The soft background, which is mainly Galactic, varies by a
factor of $\sim$2 even at high Galactic latitudes (Abell~2029 is at $b
= 50\fdg5$). Thus, it would be preferable to use an in-situ
background from the observation.  Unfortunately, the $ROSAT$ PSPC
image of Abell 2029 \citep{swm98} shows that the thermal X-ray
emission from the cluster covers the entire S3 chip, and even extends
beyond the $Chandra$ S1 chip (the other backside-illuminated chip).
As a result, it is difficult to determine an in-situ background from
the observation.  For this absorption study, it would also be useful
to separate the background into particle and Galactic foregrounds and
cosmological X-ray background.  Although the correct background is
thus uncertain by a factor of about two, the dominant ``background''
in these measurement is actually the diffuse cluster emission from
Abell 2029, and the results presented here are affected by less than
10\% by this uncertainty in the real background.  The sense of the
variation is that, if the background is higher than we have assumed
(the likely sense of the variation), we have underestimated the
absorption by the spiral galaxy, albeit very slightly.

Figure~\ref{fig:optic_xray}b,c shows Gaussian smoothed
($\sigma$=1\arcsec) images in the soft and hard X-ray bands of the
region around the spiral galaxy UZC J151054, both corrected for
exposure and background.  The emission gradient seen from north to
south in the image is due to the thermal ICM of the core of the
cluster Abell 2029. Detailed X-ray images of the galaxy cluster are
presented in Paper II.  The X-ray image was compared to the Two Micron
All Sky Survey \citep[2MASS;][]{Cut+01} to check the positional
accuracy.  There are three 2MASS IDs which are associated (within
1\arcsec) with X-ray sources, and comparison of positions shows no
significant shift between the two coordinate systems.

\subsection{Optical}

$B$, $V$, $R$, and $I$~band photometry has been obtained for a
different project by one of us (JMU) in collaboration with
S.~P.~Boughn (Haverford) and J.~R.~Kuhn (Hawaii).  The observations
were taken with the 0.9~m telescope on Kitt Peak National Observatory
on 1987~May~20 and 1988~May~10 ($R$), 1988~May~8 ($V$), and 1988~May~14
($B$).  The No.~3 RCA CCD was used for these observations mounted at
the f:7.5 Cassegrain focus, which resulted in square pixels of
0$\farcs$86 on a side.  The seeing was $\sim 1\farcs4$.  The
observations produced mosaics of Abell~2029 which included the
foreground spiral galaxy that we discuss in this paper \citep[see][
for details of the observing and data reduction techniques]{ubk91}.
$I$~band photometry of Abell~2029 was obtained on 1998~April~19 using
the T2KA camera mounted at the f:7.5 Cassegrain focus, which resulted
in square pixels of 0$\farcs$68 on a side.  Throughout the run the
seeing was excellent, between 0$\farcs$7 and 0$\farcs$9, which
resulted in an effective seeing of $\sim1\farcs$2 due to the available
pixel size.  The spatially-overlapping exposures were used to make a
mosaic of about 35$\arcmin$ (RA) by 58$\arcmin$ (Dec) again including
the galaxy discussed here \citep{dale}.  Figure~\ref{fig:optic_xray}a
shows our galaxy immersed in the diffuse halo that pervades the center
of Abell~2029 and extends about 850~kpc from the cluster center
\citep{ubk91}. The coordinate system of the $I$~band image of the
spiral galaxy was tied to the 2MASS position frame using five matches
in the field.

\subsection{Radio Continuum}

A $\lambda \sim 23$~cm continuum image has been obtained from a deep
HI survey of Abell~2029.  The NRAO Very Large Array (VLA) was used in
its C~configuration to cover the frequency range of 1300.7~MHz to
1335.2~MHz appropriate to the redshift range of the cluster.  Although
the frequency range precluded the observation of \HI\ at the redshift
of the foreground spiral, it yields a good continuum image of the
field, with a resolution of $\sim 16$\arcsec\ and an rms noise of
$\sim 58$ $\mu$Jy/beam.  Figure~\ref{fig:Rad_cont} shows that the
spiral galaxy is undetected in the radio continuum; we derive an
integrated flux of $36 \pm 82$ $\mu$Jy which leads to a $2 \sigma$
upper limit of 200 $\mu$Jy over the full area of the galaxy (270
arcsec$^2$).

\subsection{\HI\ Observations}

\begin{figure}
\plotone{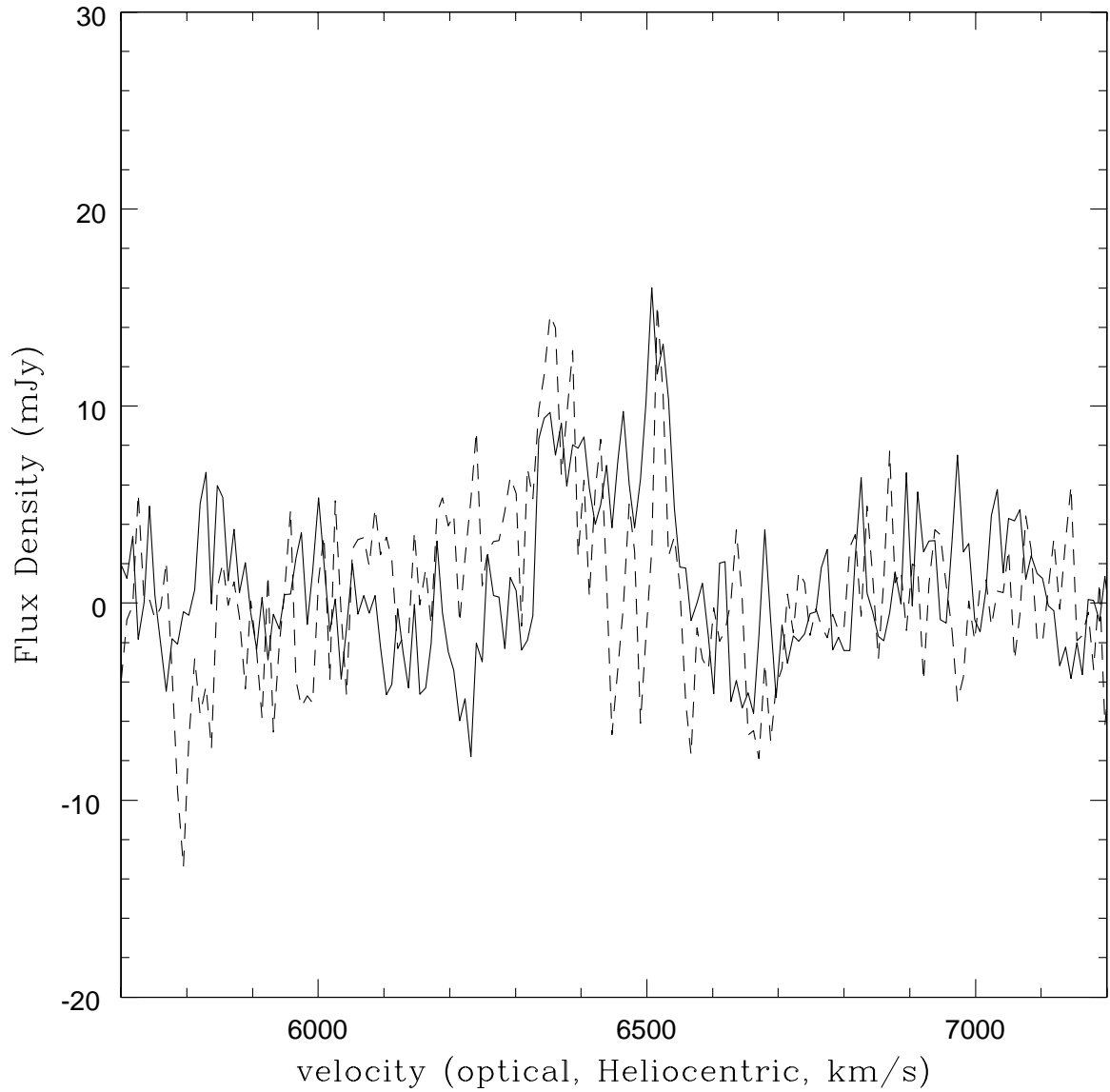}
\caption{HI spectra of UZC~J151054 obtained with the GBT.  The two curves are
different polarizations; one of the polarizations (the dashed curve)
is contaminated by ripples from out-of-band RFI.\label{fig:HI}}
\end{figure}

\HI\ observations were obtained for us by Jodie Martin (UVa) and John
Hibbard (NRAO) in two 10 minute observations with the NRAO Green Bank
Telescope (GBT). The observations were made with the Spectral
Processor using two orthogonal polarizations with bandwidths of 10~MHz
with a frequency resolution of 9.8~kHz.  The elevation was $\sim
22$~degrees which caused a fair amount of radio frequency interference
(RFI) pickup.  The data were (twice) Hanning-smoothed off-line in
order to minimize a ripple due to strong out-of-band RFI, giving a
final resolution of $\sim 8.6$~\kms.  The data were reduced using the
DISH package in AIPS++.  Even in the presence of RFI, the spectral
stability was quite adequate and we have only subtracted a linear
baseline from the data.  The calibration was referenced to 3C295 which
was observed about 12~hours after the galaxy at about the same
elevation.  We adopt a value for the flux of 3C295 of 22.46~Jy at our
observing frequency (1391~MHz) from the \citet{baars} scale which
leads to the spectrum shown in Figure~\ref{fig:HI}. From the scatter
of the data in the line free channels, we estimate the 1-$\sigma$
error per channel for the two polarizations to be 3.5~mJy (continuous
line) and 4.4~mJy (dashed line).

The average redshift is 6438~\kms\ (optical definition, heliocentric)
in good agreement with the value of $6442 \pm 3$~\kms\ listed in the
Updated Zwicky Catalog \citep{f99}, which was obtained by \citet{h97}
using the Arecibo telescope.  Correction to the frame of
the cosmic microwave background radiation yields a redshift of
c$z_{\rm cmb} = 6620$ \kms\ placing the galaxy at a distance of $\sim
90$~Mpc assuming a pure Hubble flow.  We shall use this distance to
derive physical quantities below but express those values in units of
$d_{90}$ where appropriate.

\section{Optical Properties}\label{opt_prop}

We have used the prescription in \citet{t98} to determine the
inclination and find a most-likely value of $87^\circ$, assuming an
intrinsic aspect ratio of 10:1 for the disk. The lower limit is $\ge
84^\circ$ if we assume that the galaxy is thin. We have adopted a
value of $87^\circ \pm 3^\circ$.  The Holmberg diameter is barely
reached by the observations and has a value of $\sim
24.3$~$d_{90}$~kpc.

Table~\ref{tab:Opt_phot} shows the results of the photometry.  Given
the seeing differences we only determine the photometry for the center
of the galaxy (a $6\arcsec \times 6\arcsec$ box), the bulge, the disk,
and the galaxy as a whole. To define the regions used for photometry
we took an average of the $B$, $V$, $R$, and $I$ frames to make masks
and used the same area for all bands in the photometry. The values are
corrected for foreground extinction using the prescription of
\citet*{sfd98}; but are not corrected for internal extinction (see
below).  The background was determined by first masking all sources
visible in any of the four bands, then estimating the average and
gradient due to the cD halo from the remaining pixels on either side
of the nucleus both above and below the disk. We estimate the
photometry to be accurate to $\sim 0.03$~mag given the difficulty to
estimate the local ``background'' levels in the presence of faint
cluster galaxies and the large envelope of the cD galaxy that
dominates Abell~2029. Similarly, we estimate the colors to be accurate
to $\sim$ 0.05 -- 0.1 mag.

The $V - I$ colors are consistent with those of Scd galaxies given by
\citet{fg94} with the $k$-corrections interpolated linearly to $z=
0.023$.  For example, the global color is $V - I = 0.77$ when 0.82
(1.14) is the corresponding ``average'' color of an Scd (Sbc) galaxy.
The $R - I$ color is a bit blue (0.34) where 0.53 is expected for a
typical Scd galaxy (0.77 for an Sbc), and could be affected by a
contribution from \hal\ as well.  The $B - I$ color is even bluer (1.17
when 1.37 is expected for an Scd galaxy and 1.76 for an Sbc), which
might be an indication of recent star formation.

These colors are uncorrected for internal extinction which would, of
course, make them even bluer.  There is a long-standing debate on the
optical depth of (late) spiral galaxies
\citep*[e.g.,][ and references therein]{mgh2003}.
The correction to a face-on orientation is usually
parametrized in terms of the aspect ratio as
\begin{equation}
 \Delta M_I = - \gamma_I \, \log ( a / b) \, ,
\end{equation}
where $ \Delta M_I$ is the correction to the I band magnitude, $a$ and
$b$ are the apparent semi-major and semi-minor axes of the galaxy, and
$\gamma_I$ is an empirical coefficient. Because our galaxy is about 2
magnitudes dimmer than $L^*$, $\gamma_I = 0.5$ \citep{g94} leading to
a correction of $\Delta M_I \sim -0.48$.  For comparison, we have used
the two relations offered by Tully and collaborators \citep{t98} to
compute the same correction, which in our case lead to values of
$\gamma_I = 0.3$ using the relation that depends on the absolute
magnitude, and $\gamma_I = 0.6$ using the dependence on the velocity
width.  Given the statistical character of these prescriptions, we
find their agreement quite encouraging. Adding a $k$-correction of
$-0.025$ magnitudes from \citet{fg94} leads to an $I$ band estimate of
the luminosity of $L_I = (1.1 \pm 0.2) \times 10^9 \, d_{90}^2 $~\lsun
. The expected stellar mass-to-light ratio of UZC~J151054 is $M_* /L_I
= 1.7$ in solar units \citep{deJ96} yielding an estimate of the total
stellar mass of the galaxy of $M_* \approx 1.9 \times 10^{9} \,
d_{90}^2 \, M_\odot$.

We have opted to use the average of both of the \citet{t98} methods to
similarly estimate a correction to the $B$ band photometry, which
yields a value of $\gamma_B = 0.8$.  This allows us to estimate an
absolute magnitude of $M_B = -17.79$ \citep[which includes a
$k$-correction of $-0.075$ from][]{fg94} corrected to face-on but not
for any residual effects from scattering or internal extinction.  Our
(final) estimate of the $B$ band luminosity of this galaxy is thus
$L_B \sim (2.0 \pm 0.3) \times 10^9 \, d_{90}^2 $~\lsun .  The
expected stellar mass-to-light ratio of UZC~J151054 is $M_*/L_B = 1.4$
in solar units \citep{deJ96}.  Thus, the total stellar mass of the
galaxy is expected to be $M_* \approx 2.8 \times 10^{9} \, d_{90}^2 \,
M_\odot$ in good agreement with the estimate from the $I$ band
photometry.

\section{\HI}\label{HI_prop}

The \HI\ spectrum shows an asymptotic velocity of $v_t = 108$ \kms.  Although
the \HI\ emission often extends beyond the optical emission, we derive a
lower limit to the total mass of the galaxy by estimating the mass internal
to the optical radius of 12~$d_{90}$~kpc as
$M_T \sim 3.3 \times 10^{10} d_{90} $\msun.

Integration of the \HI\ spectra yield estimates of the total fluxes of
$S_1 = 1.71 \pm 0.15 $~Jy~\kms\ and $S_2 = 1.09 \pm 0.19 $~Jy~\kms\
for polarizations one and two, respectively, where the errors are
derived from the scatter of the line-free channels and do not include
calibration errors.  As seen in Figure~\ref{fig:HI}, the second
polarization is corrupted by RFI, so that we shall ignore it
henceforth (although it does offer confirmation of the velocity extent
of the \HI ).  After our analysis was completed, we have received the
spectrum from \citet{h97} electronically.  Their measured total flux
is $S_H = 1.67 $~Jy~\kms, which we have been told should be increased
by perhaps 5\% to account for pointing errors and should be accurate
to about 15\% (M. Haynes, private communication).  Given our short
integration, we consider the agreement of the two derived total fluxes
somewhat coincidental.

Nevertheless, we are encouraged to derive an \HI\ mass from our
spectrum and arrive at a value of $M_{HI} = 3.1 \times 10^9
d_{90}^2$~\msun\ which is accurate to $\sim$ 10\%.  Thus, the galaxy
shares in the common ``dark matter'' problem with ratios of $M_{T} /
M_{HI} \sim 10 \, d_{90}^{-1}$, $M_{T} / L_B \sim 15 \, d_{90}^{-1} $
(in solar units).  The total of the stellar and \HI\ mass is
approximately $6 \times 10^{9} \, d_{90}^2 \, M_\odot$, which is still
about five times smaller than the total mass from the rotation
velocity.  Thus, the dark matter mass of the galaxy is approximately
$M_{\rm DM} \sim 2.7 \times 10^{10} \, d_{90} $\msun, out to a radius
of $\sim 12 d_{90}$~kpc.

We have also used the ``Baryonic Tully-Fisher'' relation derived by
\citet{tfd00} to estimate the distance to UZC~J151054. Given the
uncertainty that recent star formation adds to the $B$ band
determination of the stellar mass, we have used the stellar mass
derived above from the $I$ band photometry ($M_* \approx 1.9 \times
10^{9} \, d_{90}^2 \, M_\odot$). Equation 2 of \citet{tfd00} yields an
estimate of the mass in stars and gas of $M(stars+gas) = 5.3 \times
10^9$ \msun . Correcting the \HI\ mass for the He fraction yields a
value of $d_{90}=0.92$ and a luminosity distance of $D_{TF}$=83 Mpc,
in good agreement with the Hubble flow distance given the
uncertainties and the unknown peculiar velocity of UZC~J151054. Using
a value of $M_* /L_I = 1.3$ appropriate for an 8 Gyr Sc galaxy with
approximately continuous star formation \citep{deJ96}, instead of our
adopted value of 1.7 (corresponding to a 12 Gyr Sc galaxy) would lead
to $D_{TF}$=87 Mpc.

\section{X-Ray Absorption}\label{X-ray_prop}

\begin{figure}
\plotone{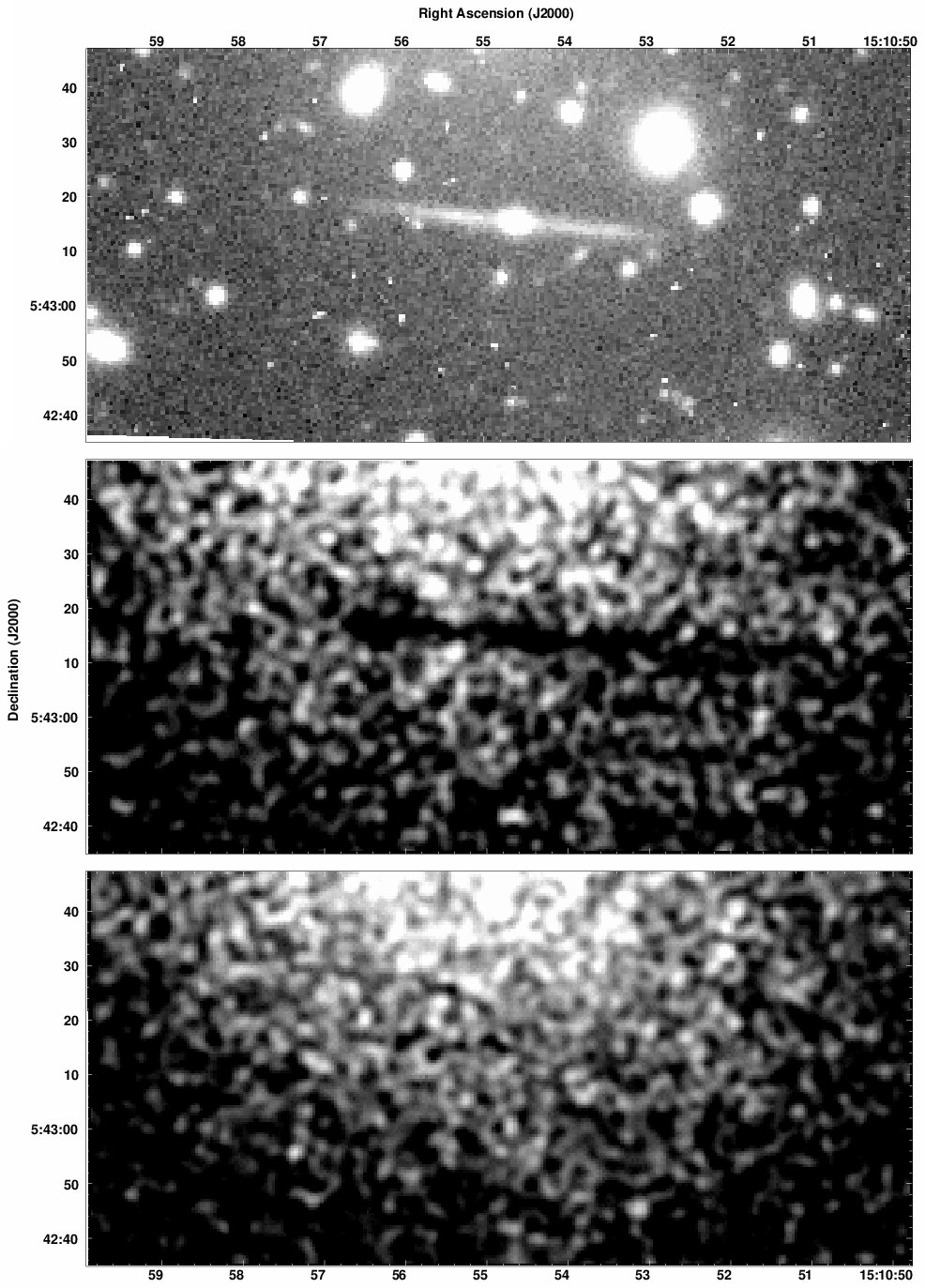}
\caption{(a) Optical I-band image of UZC J151054 taken with the KPNO
  0.9m. Lower panels show the Gaussian smoothed ($\sigma$=1\arcsec)
  $Chandra$ 0.3--1.0 keV soft X-ray image (b) and 1.0--7.0 keV hard
  X-ray image (c) of the region around the spiral galaxy. Both X-ray
  images are shown on the same intensity scale. The gradient in
  surface brightness seen north to south is due to the thermal ICM of
  the core of Abell 2029. \label{fig:optic_xray}}
\end{figure}

\begin{figure}
\plotone{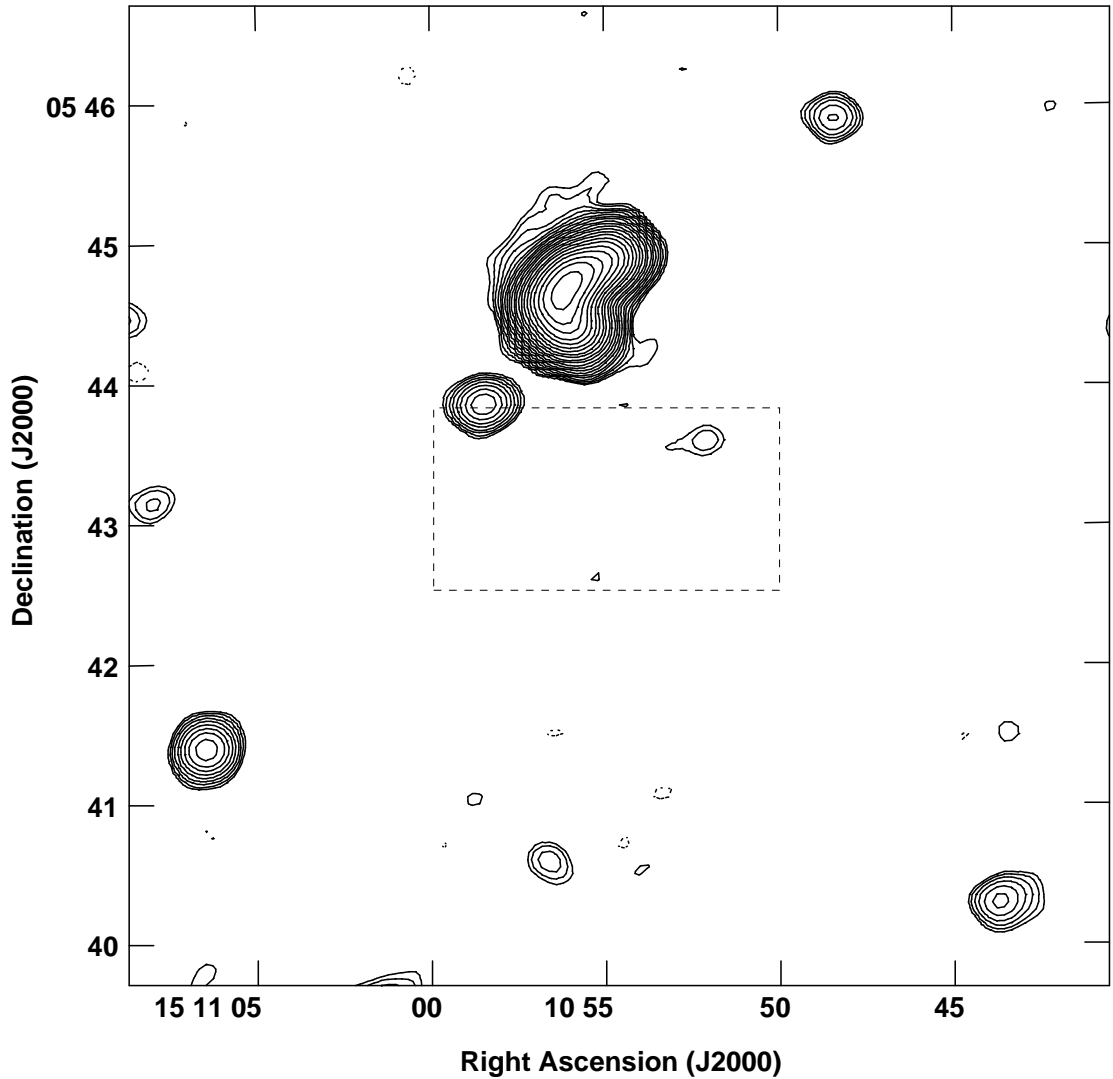}
\caption{23-cm continuum image of the 7\arcmin\ $\times$ 7\arcmin\ 
region around UZC J151054.6+054313. The dashed box corresponds to the
region shown in Figure~\ref{fig:optic_xray}. A 5.9~MHz bandpass was
used to derive this image. No continuum emission is detected from UZC
J151054.6+054313; although\ the image shows the C-shaped source
associated with the cD galaxy in Abell~2029 (convolved to the $\sim
16$\arcsec\ resolution of the image) as well as other background
emission. The contour levels are (-1.4 [absent], -1, 1, 1.4, 2, 2.8,
4, 5.6, 8, 11, 16, 22, 32, 44, 64, 88, 128, 176, 256, 352, 512, 704,
1024)$ \times 0.2$~mJy~beam$^{-1}$. (The value of 0.2~mJy~beam$^{-1}$
is the $3.5 \sigma$~level in the image).
\label{fig:Rad_cont}}
\end{figure}

Examination of Figure~\ref{fig:optic_xray} shows that the X-ray
absorption region is roughly the size of the optical disk of
UZC~J151054. Comparison of the hard (1.0--7.0 keV) and soft (0.3--1.0
keV) X-ray images indicates that the absorption is more prominent in
the soft X-ray band as expected for photoelectric absorption. The
$Chandra$ 0.3--1.0 keV image contains 82 counts in the region of
absorption, compared to 197 and 184 counts in the same size region
above and below the absorption feature. This corresponds to a $>
8\sigma$ detection of the foreground spiral galaxy absorption
feature. A careful comparison of the position angle of the optical
galaxy and the X-ray absorption feature shows that the X-ray
absorption may be slightly tilted ($\Delta PA \sim 3^\circ$) relative
to the galaxy disk (Figure~\ref{fig:optic_xray}) . Such a misalignment
could be the result of a warp in the HI disk relative to the optical
disk in UZC~J151054.

\begin{figure}
\plotone{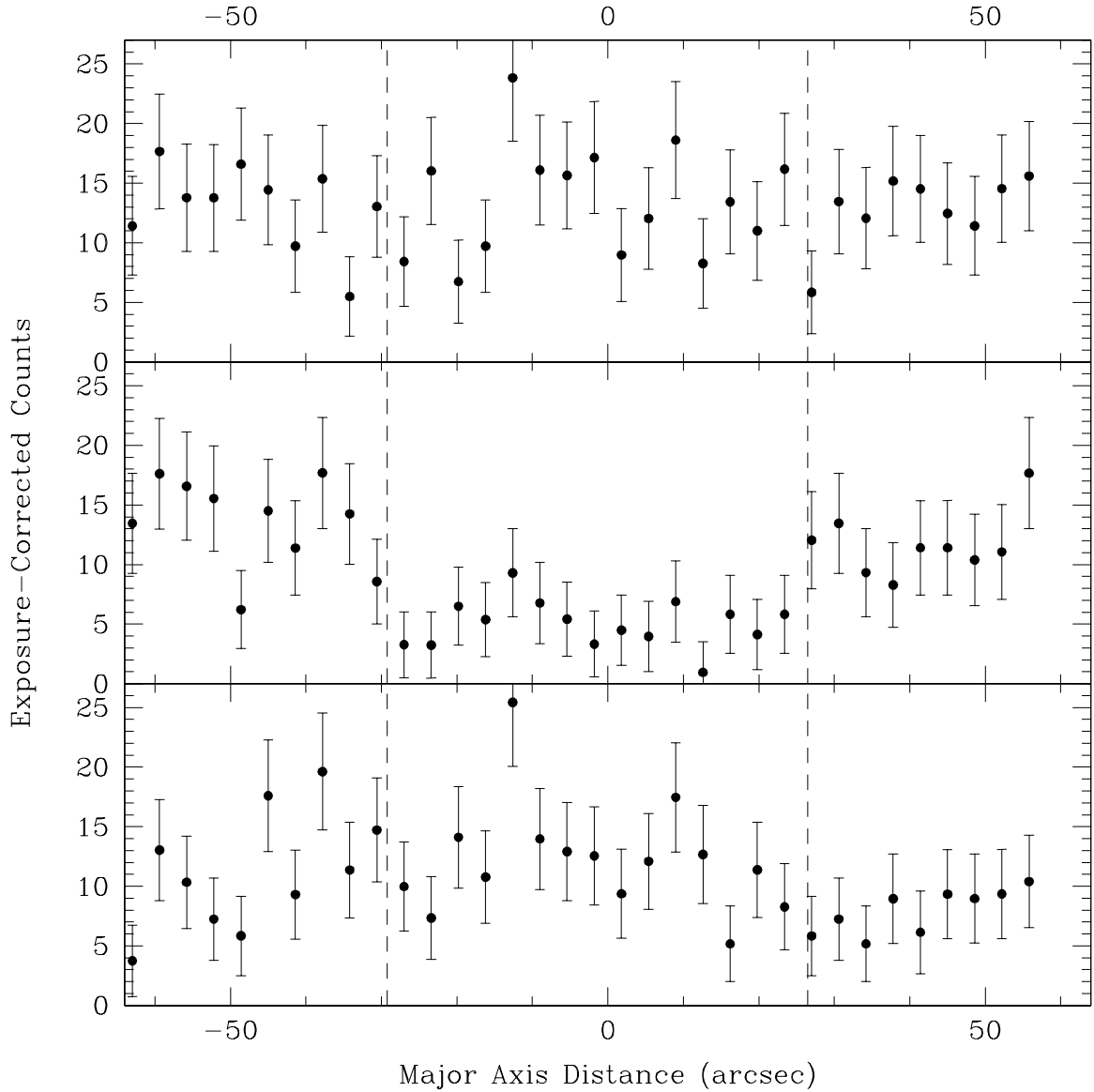}
\caption{Counts in the (unsmoothed) $Chandra$ 0.3--1.0 keV band for 34
apertures (3\farcs6 $\times$ 3\farcs6) taken across the absorption
region of the foreground spiral. The offset is measured along the
absorption east (west) as a negative (positive) distance from center
of the I band image of UZC J151054.6+054313. Middle panel shows
apertures placed east to west across the absorption region. The top
(bottom) panel shows the same apertures shifted one box up (down). The
dashed lines show the approximate extent of the I-band optical disk
emission of UZC J151054. The (background-corrected) counts in each
aperture have been corrected by the ratio of the total exposure to the
exposure in the aperture. \label{fig:counts}}
\end{figure}

The extent of the absorption region was determined by taking cuts
through the X-ray images. We used square regions of width 3\farcs6
running roughly east to west (PA=87\fdg2) along the X-ray absorption
feature and determined the counts in each region from the 0.3--1.0 keV
$Chandra$ image. The box height is approximately the size of the
optical spiral disk and the width was set small enough to allow
several boxes to be placed along the absorption feature with
sufficient counts to determine the extent of the absorption region. In
Figure~\ref{fig:counts} we show a series of exposure and background
corrected counts from 34 regions running east to west along the
absorption feature. Figure~\ref{fig:counts} also shows the results
from shifting the boxes up and down by one box height and determining
the counts both above and below the absorption feature.

\begin{figure}
\plotone{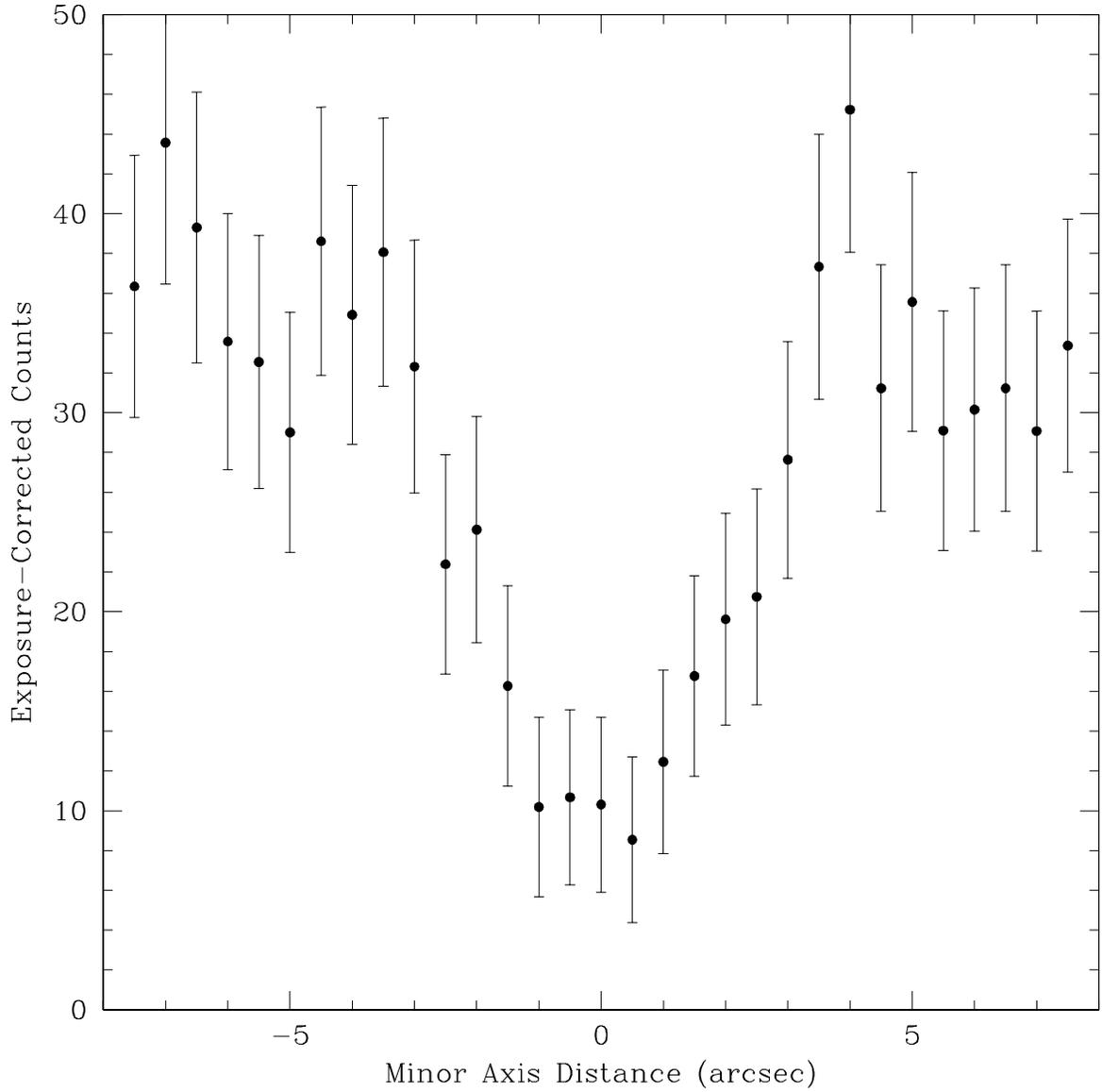}
\caption{Counts in the (unsmoothed) $Chandra$ 0.3--1.0 keV band for 31
apertures (55\arcsec\ $\times$ 0\farcs5) taken from north-south
parallel to the absorption region of the foreground spiral. The offset
is measured from the center of the absorption north (south) as a
negative (positive) distance. The (background-corrected) counts in
each aperture have been corrected by the ratio of the total exposure
to the exposure in the aperture. \label{fig:counts_perp}}
\end{figure}

In Figure~\ref{fig:counts_perp} we show a similar plot for a set of 31
regions of size 55\arcsec\ $\times$ 0\farcs5 running north-south
parallel to the disk.  Note that the PSF of the $Chandra$ ACIS-S3
detector at these positions and energies is mainly determined by the
pixel size on the detector, and the 80\% encircled energy radius is
about 0\farcs7.  Thus, the absorption from the disk is resolved
perpendicular to the disk.

The limited number of counts in the absorption region preclude any
detailed modeling of the spatial structure of the absorption.  The
full width of the absorption along the disk is about 56\arcsec\ (25
kpc).  The dashed vertical lines in Figure~\ref{fig:counts} show the
approximate I-band optical extent of the galaxy.  The width of the
absorption feature agrees fairly well with the optical width of the
galaxy.  In some ways, this might be surprising, since many spiral
galaxies have \HI\ disks which extend beyond their optical disks.  On
the other hand, the X-ray absorption at these energies is mainly due
to heavy elements (primarily oxygen), and the abundances of heavy
elements are likely to be low outside of the optical disk.  In the
vertical direction, the width of the absorption disk is about
5\arcsec\ (2.2 kpc), after making a small correction for the PSF of
the $Chandra$ ACIS-S3.  The size and strength of the X-ray absorption
can also be characterized by the equivalent width, or the width of a
region with complete absorption and the same reduction in flux as that
of the spiral galaxy absorption feature.  Using the 0.3--1.0 keV data,
we determine an equivalent width of $\sim$ 34\arcsec\ (15 kpc) along
the disk, and $\sim$ 3\arcsec\ (1.3 kpc) perpendicular to it.  These
values are smaller than the full width of the feature because the
absorption is not complete.

To analyze in more detail the absorption due to the spiral disk, we
extracted spectra from the feature and a larger surrounding
``background'' area.  The spectra were binned to a minimum of 25
cts/bin to provide good statistics. The region of the absorption
feature and the surrounding thermal ICM were fit in XSPEC with a WABS
* ZWABS * MEKAL model after applying a correction for the quantum
efficiency degradation. The MEKAL model represents the background
cluster emission, while the ZWABS is the absorption from the
foreground spiral.  The Galactic column in WABS was set to $3.14
\times 10^{20}$ cm$^{-2}$ as determined from
\citet{dl90}\footnote{Typical variations in the Galactic column
determined from \citet{dl90} around the location of UZC~J151054
($b=50\fdg5$) are less than 10\%. We also allowed the Galactic column
to be free in the fits and found a best fit value lower than the
Galactic value ($N_H$ = $1.3 \times 10^{20}$ cm$^{-2}$).  This low
Galactic hydrogen column density is consistent with that found by
\citet{lewis2002} for the outer regions of the cluster. Using the
lower Galactic column for the fits results in less than a 5\% change
in the excess absorption column due to the foreground spiral galaxy.}.
In the fit to the absorption feature, the temperature and metallicity
for the cluster emission were fixed at the value determined by the
surrounding ``background'' region, but the excess absorption was
allowed to vary. The surrounding thermal ICM is best fit with a
temperature of 7.1 keV ($\chi^2$/d.o.f.\ = 141/175), slightly lower
than the deprojected temperature found by \citet{lewis2002}.  The
model fits in XSPEC were taken for the energy range of 0.5--10.0
keV. We extended the model fits down to 0.3 keV and found consistent
results although there were significantly larger residuals at the
lowest energies in the fits of the surrounding region.

\begin{figure}
\plotone{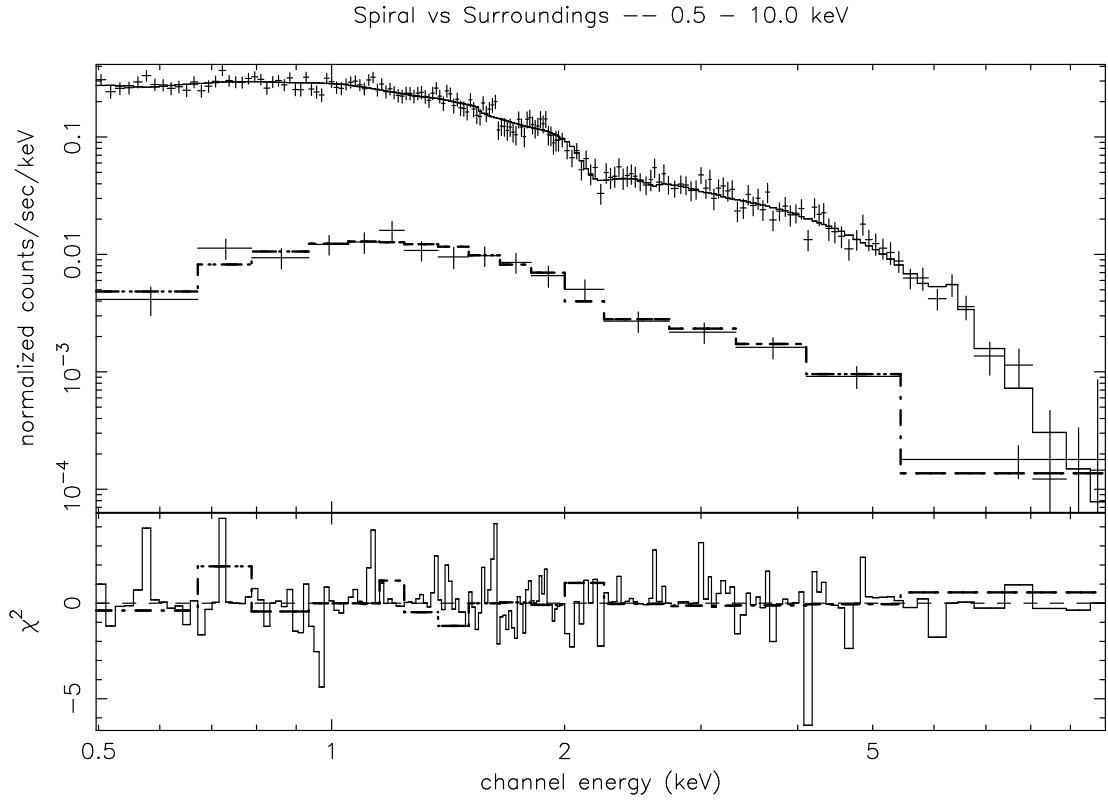}
\caption{Upper panel is the X-ray spectrum of the entire absorption
feature (lower spectrum) and the surrounding ``background'' A2029
cluster (upper spectrum). For each, the error bars give the measured
data, while the histogram is the best-fit model spectrum. The
lower panel gives the residuals to the fit in terms of their
contribution to $\chi^2$, multiplied by the sign of the
residual. \label{fig:Xspec}}
\end{figure}

The best fit model ($\chi^2$/d.o.f.\ = 152/193) to the X-ray spectrum
(Figure~\ref{fig:Xspec}) gives an excess absorbing column (with 90\%
confidence interval) of $( 2.0 \pm 0.4) \times 10^{21}$ cm$^{-2}$.  At
these columns, almost all of the absorption is due to heavy elements,
particularly oxygen.  Thus, this measurement is most directly
interpretable as a measurement of the column density of oxygen (in any
form containing K-shell electrons) of $N_O = (1.7 \pm 0.3) \times
10^{18}$ cm$^{-2}$ \citep[assuming the metallicity of][]{angr}.  The
column density of hydrogen can thus be written as $( 2.0 \pm 0.4)
\times 10^{21} \, [(O/H)/(O/H)_\odot]^{-1} $ cm$^{-2}$, where
$(O/H)/(O/H)_\odot$ is the oxygen abundance in units of the solar
abundance.  This corresponds to a mass of $M_H = (6.2 \pm 1.2) \times
10^8 \, [(O/H)/(O/H)_\odot]^{-1} \, d_{90}^2 \, $ \msun, assuming that
the absorber is spread uniformly over the 55\farcs6 by 3\farcs6 region
covered by the disk of the galaxy.  Of course, it is likely that the
interstellar gas and dust in UZC~J151054 are concentrated to the
center of the galaxy and the middle of the disk, and that the
absorption is not uniform.  Thus, it is likely that the column density
and derived mass are really lower limits to the total amount of
absorbing material.  We note that the hydrogen mass determined from
the X-ray absorption (assuming a uniform absorber and solar
abundances) is about a factor of 5 smaller than the atomic hydrogen
mass from the 21 cm observations.  This may be due to a non-uniform
distribution of the absorber, or due to a lower than solar metallicity
in the interstellar medium (ISM) in UZC~J151054.

The X-ray absorption feature appears to be limited vertically to the
approximate width of the optical disk in UZC J151054.
In order to limit the absorption due to material in the halo of
this galaxy, we also extracted the spectra from regions which were
55\farcs6 by 3\farcs6 but shifted by 3\farcs6 above and below the disk.
The spectra of the two regions above and below the disk were combined, and
fit in the same way as the absorption from the disk.
The best-fit model had an excess absorption of
$1.3^{+2.7}_{-1.3} \times 10^{20}$ cm$^{-2}$ ($\chi^2$/d.o.f.\ = 173/217).
Since the excess is consistent with zero at the 90\% confidence level, we
interpret this measurement as a 90\% confidence upper limit of
$\Delta N_H < 4.0  \times 10^{20}$ cm$^{-2}$.

\section{Discussion}\label{dis}

We have found a deep X-ray absorption feature due to a foreground
spiral galaxy seen in projection against the core of the galaxy
cluster Abell 2029.  Based on the 0.3--1.0 keV $Chandra$ data, the
X-ray deficit associated with the spiral disk of UZC J151054 is
detected at $>$ 8$\sigma$ significance.  The X-ray absorption is
strongest at low energies, as expected for photoelectric absorption.
From the X-ray spectrum of the absorption feature, we find an
absorbing column of $( 2.0 \pm 0.4) \times 10^{21}$ cm$^{-2}$,
assuming solar abundances.  This corresponds to a mass of $ (6.2 \pm
1.2) \times 10^8 \, M_\odot$ of hydrogen in the spiral disk, if the
absorber is uniformly distributed over the region of the spiral disk.
If the absorber is not uniform, as seems more likely, the required
mass of the absorber is higher.

An analysis of the optical data shows that the galaxy colors are
consistent with those of an Scd galaxy, while the axial ratio gives an
inclination of $87^\circ \pm 3^\circ$. We fail to detect radio
continuum emission from UZC J151054.6+054313 at a reasonably low
level.  Our limit of $S(1317\, {\rm MHz}) \leq 200\, \mu$Jy is
consistent with the lack of significant emission in the IR; the galaxy
is barely detectable in the 2MASS survey images \citep{Cut+01}.

The detected X-ray absorption is smaller than that expected from the
amount of neutral hydrogen detected in 21-cm emission from the disk of
UZC~J151054.  This may indicate that the ISM in UZC~J151054 is not
uniformly distributed over the absorption region, or that the
metallicity is low.  The X-ray absorption measurements also provide an
upper limit on any additional diffuse gas or dust in this galaxy.  We
note that X-ray absorption is relatively insensitive to the physical
state of the diffuse material.  Any form of gas or dust at
temperatures below $\la 10^6$ K would produce absorption.  At the
columns of interest here, most of the absorption is due to the K-shell
electrons in oxygen.  The measurements directly give the column
density of oxygen, rather than hydrogen.  Thus, the limits of the mass
of the absorbing material depend inversely on the abundance of oxygen
relative to hydrogen.

The mass of the absorbing material we detect, $M_H = (6.2 \pm 1.2)
\times 10^8 d_{90}^2 \, $ \msun, is much less than the total mass of
the galaxy or the dark matter mass, $M_{\rm DM} \sim 2.7 \times
10^{10} \, d_{90} $\msun.  Thus, our measurements limit the
possibility that the dark matter is diffuse baryonic gas.  If the dark
matter is in the form of a roughly spherical halo, then our limits on
the absorption outside the disk yield a limit on the mass of $9 \times
10^9 \, M_\odot$, which is about a factor of three smaller than the
dark matter mass.  If the dark matter is in the disk and is uniformly
distributed, then the limit on its mass is certainly less than the
total absorbing mass determined for the disk of $6 \times 10^8 \,
M_\odot$.  Thus, the dark matter cannot be diffuse gas, unless it
either has very low metal abundances or is very inhomogeneous in its
distribution.

In recent years, there have been a number of suggestions that dark
matter in galaxies is indeed baryonic, and is due to dense clouds of
gas \citep*[e.g.,][]{pcm94,os96}.  This gas might either be located in
a thick disk \citep{pcm94} or in a more spherical halo \citep{os96}.
However, in either model, the clouds of gas are rather small and
dense, and have a small covering factor.  Thus, the maximum optical
depth for absorption across the galaxy is limited by this covering
factor.  As a result, our measurements for UZC~J151054 do not refute
or strongly constrain these theories.

Finally, we note that a much longer (80 ksec) {\it Chandra}
observation of the center of Abell~2029 is planned for Cycle~5.
Among other aims, this observation should allow the absorption feature
due to UZC~J151054 to be studied in more detail.  It would also be
useful to have a 21 cm line image of the galaxy to compare the
emission line with absorption from the interstellar medium.

\acknowledgments

We thank Renzo Sancisi for a very interesting conversation about
baryonic dark matter in galaxies.  We thank Steve Boughn for useful
comments on $k$-corrections, Riccardo Giovanelli and Mort Roberts for
their insights on internal extinction in late-spiral galaxies. We also
thank Jodie Martin and John Hibbard for donating the 20~minute
observation with the GBT and collecting the data for us as well as for
the initial reduction of the data with the DISH package. Martha Haynes
kindly sent us a digitized copy of the Arecibo spectrum of UZC
J151054.6+054313.  Support for this work was provided by the National
Aeronautics and Space Administration through {\it Chandra}\/ Award
Numbers GO2-3159X and GO2-3160X, issued by the {\it Chandra}\/ X-ray
Observatory Center, which is operated by the Smithsonian Astrophysical
Observatory for and on behalf of NASA under contract
NAS8-39073. Support for E.\ L.\ B.\ was provided by NASA through the
{\it Chandra} Fellowship Program, grant award number PF1-20017, under
NASA contract number NAS8-39073. We have used the software packages
AIPS and DISH/AIPS++ of the NRAO. The National Radio Astronomy
Observatory is operated by Associated Universities, Inc., under
cooperative agreement with the National Science Foundation.  This
publication makes use of data products from the Two Micron All Sky
Survey, which is a joint project of the University of Massachusetts
and the Infrared Processing and Analysis Center/California Institute
of Technology, funded by the National Aeronautics and Space
Administration and the National Science Foundation. This research has
made use of the NASA/IPAC Extragalactic Database (NED) which is
operated by the Jet Propulsion Laboratory, California Institute of
Technology, under contract with the National Aeronautics and Space
Administration.

\clearpage

\clearpage 

%

\clearpage


\clearpage

\clearpage

\clearpage

\clearpage

%

\scriptsize
\begin {deluxetable}{lccccc}
\tablewidth{4truein}
\tablecaption{Optical Photometry}
\tablehead{
\colhead{Region}& \colhead{Area} & \colhead{$B$} & \colhead{$V$} & \colhead{$R$} & \colhead{$I$}
\\
\colhead{} & \colhead{(arcsec$^2$)} & \colhead{(mag)} & \colhead{(mag)} & \colhead{(mag)} & \colhead{(mag)}
}
\startdata
Center &  36 & 18.70 & 18.21 & 17.70 & 17.31\\ 
Bulge  &  58 & 18.55 & 18.05 & 17.59 & 17.20\\
Disk   & 212 & 18.60 & 18.31 & 17.92 & 17.66\\
Galaxy & 270 & 17.82 & 17.42 & 16.99 & 16.65\\
\enddata
\label{tab:Opt_phot}
\end{deluxetable}
\normalsize

\end{document}